\documentstyle[prl,aps,epsf,amssymb,twocolumn]{revtex}

\begin{document}

% \draft command makes pacs numbers print
\draft

\title{A quantum point contact for neutral atoms}
\author{J.\ H.\ Thywissen,\cite{byline} R.\ M.\ Westervelt,\cite{byline2} 
and M.\ Prentiss}
\address{Department of Physics, Harvard University, Cambridge,
Massachusetts 02138, USA}
\date{July 19, 1999}
\maketitle
\begin{abstract}
%
%         1         2         3         4         5         6        
%23456789 123456789 123456789 123456789 123456789 123456789 123456789
%
We show that the conductance of atoms through a tightly confining
waveguide constriction is quantized in units of $\lambda_{\rm
dB}^2/\pi$, where $\lambda_{dB \rm}$ is the de Broglie wavelength of
the incident atoms.
Such a constriction forms the atom analogue of an electron quantum
point contact and is an example of quantum transport of neutral atoms
in an aperiodic system.
We present a practical constriction geometry that can be realized
using a microfabricated magnetic waveguide, and discuss how a pair of
such constrictions can be used to study the quantum statistics of
weakly interacting gases in small traps.
\end{abstract}
% insert suggested PACS numbers in braces on next line
 % end of PACS codes
%
%%%%%%%%%%%%%%%%%%%%%%%%%%%%%%%%% body %%%%%%%%%%%%%%%%%%%%%%%%%%%%%%%%%
%

%PARAGRAPH ONE: [intro]
%
Quantum transport, in which the center-of-mass motion of particles is
dominated by quantum mechanical effects, has been observed in both
electron and neutral-atom systems.
Pioneering experiments demonstrated quantum transport in periodic
structures. For example, Bloch oscillations and Wannier-Stark ladders
were observed in the conduction of electrons through superlattices
\cite{superlattices} with an applied electric field, as well as in the
transport of neutral atoms through accelerating optical lattices
\cite{lattices,qt}.
Further work with neutral atoms in optical lattices has utilized their
slower time scales (kHz instead of THz) and longer coherence lengths to
observe a clear signature of dynamical Bloch band suppression
\cite{dynamic}, an effect originally predicted for but not yet
observed in electron transport \cite{dynpred}.
%

%PARAGRAPH TWO:
Quantum transport also occurs in aperiodic systems. For example, a
quantum point contact (QPC) is a single constriction through which
the conductance is always an integer multiple of some base conductance.
The quantization of electron conductance in multiples of $2 e^2 / h$,
where $e$ is the charge of the electron and $h$ is Planck's constant,
is observed through channels whose width is comparable to the Fermi
wavelength $\lambda_F$. Experimental realizations of a QPC include a
sharp metallic tip contacting a surface \cite{qpc_wires} and an
electrostatic constriction in a two-dimensional electron gas
\cite{QPC,beenakker}.
Electron QPC's have length-to-width ratios less than 10 because
phase-coherent transport requires that channels must be shorter than the
mean free path between scattering events, $\ell_{\rm mfp \rm}$.
Geometric constraints are the limiting factor in the accuracy
of quantization in an electron QPC \cite{accuracy}.
%

%PARAGRAPH THREE:
In this Letter, we present an experimentally realizable system that
forms a QPC for neutral atoms --- a constriction whose ground state
width $b_o$ is comparable to $\lambda_{\rm dB}/2\pi$, where
$\lambda_{\rm dB}$ is the de Broglie wavelength of the atoms.
The ``conductance'', as defined below, through a QPC for atoms is
quantized in integer multiples of $\lambda_{\rm dB}^2/ \pi$.
The absence of frozen-in disorder, the low rate of inter-atomic
scattering ($\ell_{mfp \rm} \sim 1$ m), and the availability of nearly
monochromatic matter waves with de Broglie wavelengths $\lambda_{\rm
dB} \sim 50$ nm \cite{atomlaser,hagley} offer the possibility of
conductance quantization through a cylindrical constriction with a
length-to-width ratio of $\sim 10^5$.
This new regime is interesting because deleterious effects such as
reflection and inter-mode nonadiabatic transitions are minimized
\cite{reflection}, allowing for accuracy of conductance
quantization limited only by finite-temperature effects.
Furthermore, the observation of conductance quantization at new
energy and length scales is of inherent interest.

%PARAGRAPH FOUR: [multiparticle effects]
If a QPC for neutral atoms were realized, it would provide excellent
opportunities for exploring the physics of small ensembles of weakly
interacting gases.
For instance, the transmission through a series of two QPC's would
depend on the energetics of atoms confined in the trap between the two
constrictions.
The physics of such a ``quantum dot'' for atoms is
fundamentally different from that of electrons, since the 
Coulombic charging energy that dominates the energetics of 
an electron quantum dot \cite{mesoscopics} is absent for
neutral particles.
The quantum statistics of neutral atoms energetically restricted
to sub-dimensional spaces has already aroused theoretical interest
in novel effects such as Fermionization \cite{tonks} and the formation
of a Luttinger liquid \cite{luttinger}. 
%

\begin{figure}
\begin{center}
\leavevmode
\epsfxsize=0.45\textwidth
\epsffile{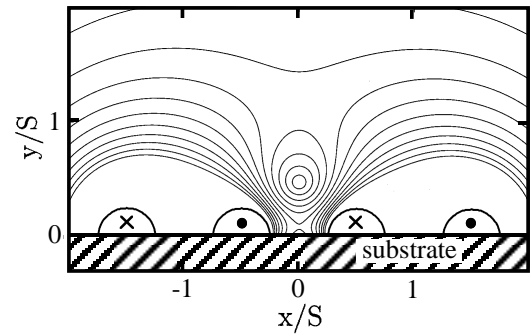}
\end{center}
\caption{Magnetic field contours above a micro-electromagnet
waveguide. Four parallel wires, separated by a distance $S$ and with
anti-parallel current flow (marked ``$\cdot$'' for $\bbox{+z}$ and
``$\times$'' for $\bbox{-z}$), are mounted on a substrate (crosshatched),
which serves both to support the wires mechanically and to dissipate the
heat produced.  A potential minimum is formed above the wires and can be
used to guide atoms in the out-of-plane direction $\bbox{z}$. Twelve
contours, equally spaced by $B_o/4$, are shown, where $B_o = \mu_o I/2 \pi
S$ and $\pm I$ ($\pm 2I$) is the current in the inner (outer) wire pair.}
\label{fig:1}
\end{figure}

%PARAGRAPH FIVE: [waveguide]
Recently, several waveguides have been proposed
\cite{hinds-2D,schmiedmayer,magwg} 
whose confinement may be strong enough to meet the constraint $b_o
\lesssim \lambda_{dB \rm}/2 \pi$ for longitudinally free atoms.
In this work, we will focus on the example of a surface-mounted
four-wire electromagnet waveguide for atoms \cite{magwg} (see Fig.\
\ref{fig:1}) which exploits recent advances in microfabricated atom
optics \cite{libbrecht,microEM}. A neutral atom with a magnetic
quantum number $m$ experiences a linear Zeeman potential $U(\bbox{r})
= \mu_{B} g m \left| \bbox{B}(\bbox{r}) \right|$, where $\mu_{B}$ is
the Bohr magneton, $g$ is the Land\'{e} g factor, and
$\bbox{B}(\bbox{r})$ is the magnetic field at $\bbox{r}$. Atoms with
$m > 0$ are transversely confined \mbox{near} the minimum in field
magnitude shown in Fig.\ \ref{fig:1}; however, they are free to move
in the $\bbox{z}$ direction, parallel to the wires.
Non-adiabatic changes in $m$ near the field minimum can be
exponentially suppressed with a holding field  $B_h$ applied
in the axial direction $\bbox{z}$ \cite{sukumar}.
Near the guide center, the potential forms a cylindrically symmetric
two-dimensional simple harmonic oscillator with classical
oscillation frequency $\omega = \left[ \mu_{B} g m (2 \mu_0 I/ \pi
S^2)^2/M B_h \right]^{1/2}$, where $\mu_0$ is the permeability of free space, $I$ is the
inner wire current, $2I$ is the outer wire current,
$S$ is the center-to-center wire spacing, and $M$ is the
mass of the atoms.
Sodium ($^{23}$Na) in the $|F=1, m_{F} = +1>$ state would have a classical
oscillation frequency of $\omega = 2\pi \times 3.3$ MHz and a root mean
squared (RMS) ground state width $b = \sqrt{\hbar/2 M \omega} = 8.1$ nm in a
waveguide with $S=1$ $\mu$m and $I=0.1$ A.
The fabrication of electromagnet waveguides of this size scale and current
capacity has been demonstrated \cite{microEM}.
%
%

\begin{figure}
\begin{center}
\leavevmode
\epsfxsize=0.45\textwidth
\epsffile{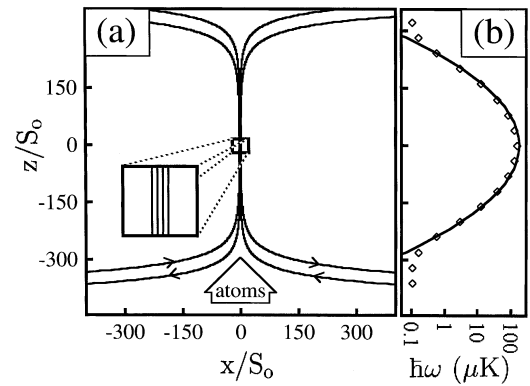}
\end{center}
\caption{{\bf (a)} Top-down view of a waveguide wire geometry which creates
a quantum point contact for atoms. The direction of current flow is
indicated on the wires (solid lines). A constriction with $\ell = 100
S_o$ is shown.
{\bf (b)} Level spacing $\hbar \omega$ (in $\mu$K) of transverse oscillator
states versus axial distance $z$. Points ($\diamond$) are based on
numerical calculations of the field curvature at each $z$ above the wire
configuration shown in (a); the line is based on the parallel-wire scaling
$S(z)^{-2}$. Both calculations assume Na atoms in the $|F=1, m_{F} =
+1\rangle $ state, $S_o = 1$ $\mu$m, $I = 200$ mA, and $B_h = 35$ G.}
\label{fig:2}
\end{figure}

%PARAGRAPH SIX: [constriction]
A constriction in the waveguide potential can be created by
contracting the spacing between the wires of the waveguide. The
constriction strength can be tuned dynamically by changing the current
in the wires. Fig.\ \ref{fig:2}a shows a top-down view of a
constriction whose wire spacing $S(z)$ is smoothly varied as
\begin{equation}
S(z) = S_o \exp{ \left( \frac{z^2}{2 \ell^2} \right) }, \label{eq:shape}
\end{equation}
where $S_o$ is the spacing at $z=0$, and $\ell$ is the characteristic
channel length.  Assuming the wires are nearly parallel, the guide
width, depth, oscillation frequency, and curvature scale as $S(z)$,
$S(z)^{-1}$, $S(z)^{-2}$, and $S(z)^{-4}$, respectively. For $\ell =
100 S_o$, field calculations above this curved-wire geometry show that
the parallel-wire approximation is valid for $|z| \lesssim 3\ell$,
allowing for a well-defined waveguide potential over a factor of more
than $10^3$ in level spacing (see Fig.\ \ref{fig:2}b). Our particular
choice of $S(z)$ is somewhat arbitrary but prescribes one way in which
wires can form a smooth, constricting waveguide as well as run to
contact pads (necessary to connect the wires to a power supply) far
enough from the channel ($\gg \ell$) that their geometry is
unimportant. The total ``footprint'' of this device (not including
contact pads) is approximately $10 \ell \times 10 \ell$, or about 1
mm$^2$, for $S_o = 1$ $\mu$m and $\ell = 100 S_o$.
%

%PARAGRAPH SEVEN: [simulation]
Atoms approach the constriction from the $\bbox{-z}$ direction, as shown in
Fig.\ \ref{fig:2}a. We calculate the propagation of the atom waves through
the constriction by solving the time-dependent Schr\"odinger
equation in three spatial dimensions. It is important to note that the
nature of quantum transport requires fully quantum-mechanical calculations,
even for the longitudinal degree of freedom within the waveguide. The
Hamiltonian for an atom near the axis of the four-wire waveguide described
by Eq.\ (\ref{eq:shape}) is
\begin{equation}
\hat{H}_{QPC} = \frac{\bbox{\hat{p}}^2}{2M} + \frac{1}{2} M \omega_o^2 e^
{- 2 \hat{z}^2 / \ell^2} (\hat{x}^2 + \hat{y}^2),
\end{equation}
where $\hat{ }$ denotes an operator, $\omega_o$ is the transverse
oscillation frequency at $z=0$, and we have assumed the parallel-wire
scaling of field curvature, $S(z)^{-4}$.
Since a direct numerical integration approach is computationally
prohibitive, we developed a model that neglects non-adiabatic
propagation at the entrance and exit of the channel. The waveguide
potential is truncated at $z = \pm z_T$, the planes between which
atoms can propagate adiabatically in the waveguide, and the
wavefunction amplitude $\psi$ and its normal derivative $\partial
\psi/\partial z$ are matched between plane-wave states ($|z| > z_T$)
and the modes of the waveguide ($|z| < z_T$). 
We found that, for $\ell \gtrsim 10 b_o$, a two-dimensional version of
the model could reproduce the transmissions and spatial output
distributions of a two dimensional split-operator FFT integration of
$\hat{H}_{QPC}(\hat{x},\hat{z})$ with the full waveguide potential.
This agreement gave us confidence in our three-dimensional model of
atom propagation through the constriction.

%PARAGRAPH EIGHT: [source]
The cross-section for an incident atomic plane wave to be transmitted
through a constriction is dependent on the plane-wave energy $E_I$ and
incident angle.
However, if the RMS angular spread of incident plane waves $\sigma$ is
much greater than the RMS acceptance angle $\alpha \sim \left[
ln(\ell/b_o) b_o^2/\ell^2 \right]^{1/4}$, we can integrate over all
solid angles and define a ``conductance'' $\Phi$ dependent only on
parameters of the constriction and the kinetic energy $E_I$ of the
incident atoms:
\begin{equation}
\Phi (E_I) = \frac{F}{J_o f(0,0)}, \label{eq:Seff}
\end{equation}
where $F$ is the total flux of atoms (in s$^{-1}$) transmitted though
the constriction and $J_o f(0,0)$ is the incident on-axis brightness
(in cm$^{-2}$s$^{-1}$). The transverse momentum distribution $f(k_x,
k_y)$ is defined as follows: in the plane wave basis
$\{|\bbox{k}\rangle\}$, we consider a density distribution of atoms on
the energy shell $a(\bbox{k}) d\bbox{k} = (C / k_z^o) \delta\left[k_z
- k_z^o \right] f(k_x, k_y) d\bbox{k}$, where $C = \hbar J_o / 2 \pi
E_I$, $\hbar k_z^o = \left[ 2 M E_I - \hbar^2 (k_x^2 +
k_y^2)\right]^{1/2}$, and $f(k_x, k_y)$ is normalized such that the
incident flux density $J_o = \int d\bbox{k} a(\bbox{k}) \hbar k_z/M$.
When applied to the diffusion of an isotropic gas ($f=1$) through a
hole in a thin wall, $\Phi$ is equal to the area of the hole; for a
channel with a small acceptance angle, $\alpha \ll
\sigma$, $\Phi$ is the effective area at the narrowest cross-section
of the channel.
We consider a distribution of incident energies $g(E_I)$ with a RMS
spread $\Delta E$, centered about $\overline{E_I}$. As an example, the
$^{23}$Na source described in Ref.\ \ref{ref:hagley} has a
monochromaticity $\overline{E_I}/\Delta E \approx 50$ for atoms
traveling at $30$ cm$/$s, or $\lambda_{\rm dB} = 50$ nm. To meet the
constraint $\sigma \gg \alpha$ \cite{constraint}, such a source can be
reflected off of a diffuser \cite{light}, such as the de-magnetized
magnetic tape described in Ref.\ \ref{ref:roach}. Assuming the spatial
density of atoms is preserved during propagation \cite{guidedmot},
such a source can have a flux density of $J_o \approx 2 \times
10^{10}$ cm$^{-2}$s$^{-1}$.
 
\begin{figure}
\begin{center}
\leavevmode
\epsfxsize=0.45\textwidth
\epsffile{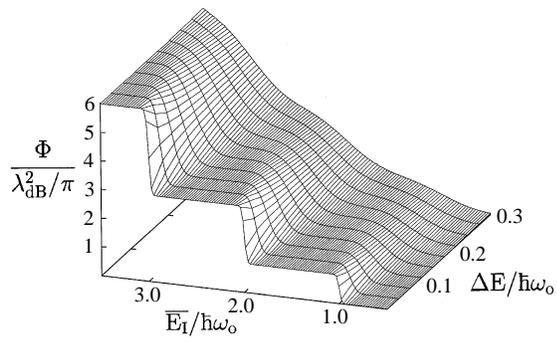}
\end{center}
\caption{Conductance $\Phi$ through a quantum point contact, as a function of
average incident energy $\overline{E_I}$ and energy spread $\Delta E$.  $\Phi$
is plotted in terms of the quantized unit of conductance, $\lambda_{dB
\rm}^2/\pi$, and $\overline{E_I}$ and $k_B T$ are plotted in terms of
$\hbar \omega_o$, the level spacing at the narrowest point of the
constriction. The lowest $\Delta E$ shown, $0.02 \hbar \omega_o$,
corresponds to the example for $^{23}$Na discussed in the text.}
\label{fig:3}
\end{figure}

%PARAGRAPH NINE: [results]
The quantized conductance for atoms is shown in Fig.\ \ref{fig:3} and
is the central result of this Letter. Conductance $\Phi/(\lambda_{dB
\rm}^2/\pi)$ is shown as a function of mean energy $\overline{E_I}/\hbar
\omega_o$ and energy spread $\Delta E/\hbar \omega_o$. In the limits
$\hbar \omega_o \gg \Delta E$ and $\ell \gg b_o$, one can show analytically
that the conductance is $\Phi = (\lambda_{dB \rm}^2/\pi) N$, where $N$ is the
number of modes above cutoff at $z=0$.
The ``staircase'' of $\Phi$ versus $\overline{E_I}/\hbar \omega_o$ is a
vivid example of quantum transport, as it demonstrates the quantum
mechanical nature of the center-of-mass motion. For all of Fig.\
\ref{fig:3} we have assumed $\ell = 10^{3} S_o \approx 10^5 b_o$; in the
particular case of the Na source discussed above, and assuming $\sigma
= 25$ mrad, the first step ($\Phi = \lambda_{dB \rm}^2/ \pi$)
corresponds to a transmitted flux of $\sim$ 500 atoms s$^{-1}$,
which is a sufficient flux to measure via photoionization. 
%
%

%
%PARAGRAPH TEN: [discussion]
We can understand several features shown in Fig.\ \ref{fig:3} by
considering the adiabatic motion of atoms within the waveguide.
As atom waves propagate though the constricting waveguide, modes with
transverse oscillator states ${(n_x, n_y)}$ such that $\hbar \omega_o
(n_x + n_y + 1) - \overline{E_I} \gtrsim 2 M \hbar^2 / \ell^2$ will
contribute negligible evanescent transmission and adiabatically
reflect before $z=0$.
Steps occur when the number of allowed propagating modes changes: the
$m^{th}$ step appears at $\hbar \omega_o = \overline{E_I}/m$. Note that
this condition can also be written $b_o = \sqrt{m} \lambda_{dB \rm}/2\pi$,
demonstrating that transverse confinement on the order of $\lambda_{dB
\rm}/2 \pi$ is essential to seeing conductance steps in a QPC.
Since low-lying modes occupy a circularly symmetric part of the potential, 
the $m^{th}$ step involves $m$ degenerate modes and is $m$ times as high
as the first step.
The large aspect ratio of the atom QPC allows for a sufficiently
gentle constriction to suppress partial reflection at the entrance to
the guide, such that the sharpness of steps and flatness between them
is limited only by the spread in incident atom energies.
%
%

%PARAGRAPH ELEVEN: [correspondence]
It is interesting to compare the electron and atom QPC systems. 
If contact is made between two Fermi seas whose chemical potentials
differ by $e \Delta V < k_B T \ll E_F$, where $\Delta V$
is the applied voltage, $T$ is the temperature of the electron gas, and
$E_F$ is the Fermi energy, then the current that flows between them
will be carried by electrons with an energy spread $k_B T$ and a mean
energy $E_F$.
For a cold atom beam, the particle flow is driven by kinetics
instead of energetics. The incident kinetic energy $\overline{E_I}$
corresponds to $E_F$, and the energy spread $\Delta E \ll
\overline{E_I}$ corresponds to $k_B T$.
The quantum of conductance for both systems can be formulated in terms
of particle wavelength: the classical conductance of a point contact
of area A connecting two three-dimensional gases of electrons is $G =
(e^2 k_F^2 A)/(4 \pi^2 \hbar)$ \cite{sharvin}, such that if 
$G = N e^2/ \pi \hbar$, the effective area is $A = N \lambda_F^2/\pi$.
In order to determine the accuracy of conductance quantization, three
measurements ($F$, $J_o f(0,0)$, and $\lambda_{dB}$) are necessary for
the atom QPC instead of two measurements (current and $\Delta V$) for the
electron QPC.
The reduced number of degrees of freedom for electrons results from
their Fermi degeneracy: the net current is carried by electrons whose
incident flux density $J_o$ and wavelength $\lambda_{F}$ are functions
of $E_F$ and $\Delta V$.
As a thought experiment, the simplicity of an externally tuned $J_o$
and $\lambda_{\rm dB}$ could also be extended to neutral atoms, if two
degenerate ensembles of Fermionic atoms were connected by a QPC and
given a potential difference $\Delta U$, such as could be induced by a
uniform magnetic field applied to one reservoir.
We can redefine neutral atom conductance as $\Gamma =
F/\Delta U$, where $F$ is the transmitted atom flux, just as the electron
conductance $G$ is the ratio of electron flux (current) to potential
difference (voltage).
One can show that
\begin{equation}
\Gamma = \frac{N}{h},
\end{equation}
assuming $\Delta U < k_B T \ll E_F$, where $T$ is the temperature of
the Fermi ensembles and N is the number of modes above cutoff.

%PARAGRAPH TWELVE [extensions]
Two QPC's can form a trap between them, just as a pair of electron
QPC's form a quantum dot \cite{mesoscopics}.  For $\hbar \omega_o >
\overline{E_I}$, all modes of the QPC are below cutoff and evanescent
transmission is dominated by tunneling of atoms occupying the $(0,0)$
mode. While the quantum dot between them is energetically isolated,
atoms can still tunnel into and out of the dot. 
For cold Fermionic atoms, the Pauli exclusion principle would enable a
single atom to block transmission through the trap, just as the
charging energy of a single electron can block transmission in
electron quantum dots; such a blockade might be used to make a
single-atom transistor.
In such a single-atom blockade regime, quantum dots can also show a
suppression of shot noise below the Poissonian level \cite{mesoscopics}.
Note that spectroscopic measurement of neutral atom traps with
resolvable energy levels has been suggested previously
\cite{schmiedmayer} in analogy to spectroscopic measurement of
electron quantum dots. We emphasize that the loading and observation
of such a small trap with two or more QPC ``leads'' is a powerful
configuration for atom optics, because loading a small, isolated trap
is problematic, and because spectroscopy near the substrate is
complicated by light scattering and inaccessibility.
%

%PARAGRAPH THIRTEEN: [conclusion]
In conclusion, we show how an electromagnet waveguide could be used to
create a quantum point contact for cold neutral atoms. This device is
an example of a new physical regime, quantum transport within
microfabricated atom optics.

%%%%%%%%%%%%%%%%%%%%%%%%%%%%%%%%% acknowledgements %%%%%%%%%%%%%%%%%%%%%
%
The authors thank A.\ Barnett, N.\ Dekker, M.\ Drndi\'{c}, E.\ W.\
Hagley, K.\ S.\ Johnson, M.\ Olshanii, W.\ D.\ Phillips, M.\ G.\
Raizen, and G.\ Zabow for useful discussions. This work was supported
in part by NSF Grant Nos.\ PHY-9732449 and PHY-9876929, and by MRSEC
Grant No.\ DMR-9809363. J.\ T.\ acknowledges support from the Fannie
and John Hertz Foundation.
%
%%%%%%%%%%%%%%%%%%%%%%%%%%%%%%%%%% bibliography %%%%%%%%%%%%%%%%%%%%%%%%
%

%

\begin{references}
\bibitem{constraint}{If a source with $\sigma \lesssim \alpha$ were used,
then $\Phi$ would still have steps at the same energies, but the
plateaus in Fig.\ \ref{fig:3} would no longer be flat.}
\end{references}
\end{document}